\documentclass[global,twocolumn]{svjour}

\usepackage{times}
\usepackage{graphicx} 
\usepackage{dcolumn} 
\usepackage{bm} 

\journalname{Applied Physics B}

\begin{document}


\title{Photoassociation inside an optical dipole trap: absolute rate 
       coefficients and Franck-Condon factors}

\author{
R. Wester \inst{1}
\and S. D. Kraft \inst{1}
\and M. Mudrich \inst{1}
\thanks{Present address: Laboratoire Aim{\'e} Cotton, CNRS, 
  B{\^a}timent 505, Campus d'Orsay, 91405 Orsay, France}
\and M. U. Staudt \inst{1}
\and J. Lange \inst{1}
\and N. Vanhaecke \inst{2}
\thanks{Present address: Fritz-Haber Institut, Faradayweg 4-6,
14195 Berlin, Germany}
\and O. Dulieu \inst{2}
\and M. Weidem{\"u}ller \inst{1}
}
\institute{Physikalisches Institut, Universit{\"a}t Freiburg,
  Hermann-Herder-Stra{\ss}e 3, 79104 Freiburg, Germany
\and
 Laboratoire Aim{\'e} Cotton, CNRS, B{\^a}timent 505, Campus 
d'Orsay, 91405 Orsay, France}

\date{\today}

\maketitle


\begin{abstract}
We present quantitative measurements of the photoassociation of cesium
molecules inside a far-detuned optical dipole trap. A model of the trap
depletion dynamics is derived which allows to extract absolute
photoassociation rate coefficients for the initial single-photon
photoassociation step from measured trap-loss spectra. The sensitivity of
this approach is demonstrated by measuring the Franck-Condon modulation of the
weak photoassociation transitions into the low vibrational levels of the outer
well of the $0_g^-$ state that correlates to the $6s+6p_{3/2}$ asymptote. The
measurements are compared to theoretical predictions. In a magneto-optical
trap these transitions have previously only been observed indirectly through
ionization of ground state molecules.
\end{abstract}


\section{\label{sec:intro}Introduction}

Formation, manipulation and interactions of ultracold molecules today forms
one of the major topics in ultracold atomic and molecular physics.
Photoassociation of a cold atom pair followed by spontaneous emission to the
ground state, which has started the field of ultracold molecules
\cite{fioretti1998:prl,takekoshi1999:pra}, and coherent magnetic field sweeps
across Feshbach resonances represent the most prominent techniques to form
ultracold molecules; the latter has finally lead to the successful creation of
molecular Bose Einstein condensates
\cite{jochim2003:sci,greiner2003:nat,zwierlein2003:prl}.  To populate
low-lying molecular states that are not accessible by the Feshbach resonance
method, two-color photoassociation schemes have been investigated
\cite{nikolov2000:prl,lisdat2002:ejd}.  Coherent photoassociation of a Mott
insulator phase with exactly two atoms per lattice site into a molecular phase
has combined the two approaches employing the scheme of a laser-driven
Feshbach resonance \cite{rom2004:prl}.  The formation of
heteronuclear cold molecules composed of two different alkali atoms has
only recently been successfully achieved by photoassociation in a double
magneto-optical trap \cite{kerman2003:prl,mancini2004:prl}.

We have investigated photoassociation inside a far-detuned optical dipole
trap, because this type of trap offers the possibility to catch associated
molecules that decay into bound levels of the ground electronic state inside
the same optical dipole trap and store or manipulate them for long periods of
time. This approach will therefore facilitate collision and chemical reaction
experiments with state-selected ultracold molecules and may culminate in
coherent ultracold chemistry \cite{heinzen2000:prl}. While most
photoassociation experiments so far have employed magneto-optical traps, the
first demonstration of the photoassociation of ultracold atoms employed an
optical dipole trap \cite{miller1993:prl}. The higher densities and the
suppression of near resonant light scattering in optical dipole traps allow
for high photoassociation rates and long photoassociation times.  As a
consequence, a strong loss of trapped atoms due to photoassociation is
achieved. Furthermore, the initial atomic quantum state is not determined by
the trap, in contrast to magneto-optical or magnetic traps, but can be
controlled independently.

\begin{figure}
\includegraphics[width=8cm]{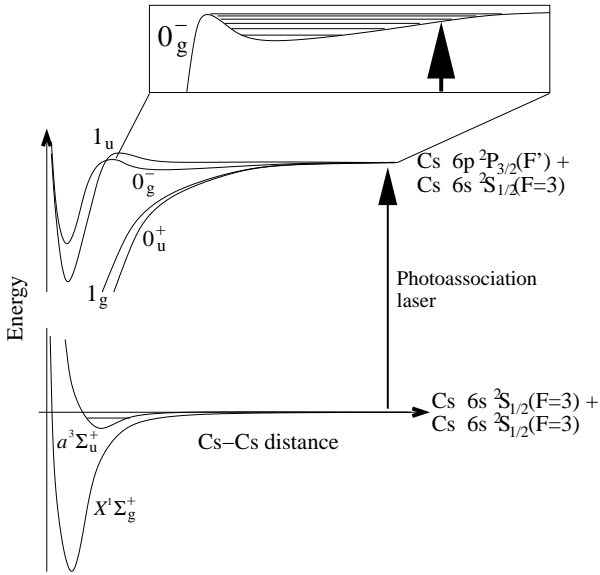}
\caption{\label{fig:photoassociation} Schematic overview of the relevant
  molecular potential curves for Cs$_2$ photoassociation into the 0$_g^-$
  outer well.}
\end{figure}

In this work we present photoassociation trap loss spectra of ultracold cesium
inside a quasi-static optical dipole trap formed by a CO$_2$ laser. We focus
on the $0_g^-(6s+6p_{3/2}(^2P_{3/2}))$ double-well state of Cs$_2$ (hereafter
referred to as $0_g^-(P_{3/2})$) which is extensively described in Ref.\
\cite{fioretti1999:ejd}. Fig.\ \ref{fig:photoassociation} schematically shows
this molecular potential curve together with the curves of other relevant
electronic states that correlate to the $6s+6p_{3/2}$ asymptote. The
$0_g^-(P_{3/2})$ state is particularly interesting, because single-photon
photoassociation allows to populate vibrational levels in its outer well that
have a high probability to decay into bound levels of the electronic ground
states. This is due to the barrier between the inner and outer well of this
potential which leads to large Franck-Condon factors for the radiative decay
\cite{fioretti1998:prl}. Also tunneling into the inner part of the potential
has been observed \cite{vatasescu2000:pra}. The highly sensitive detection of
ground state molecules by multiphoton ionization was used in a sequence of
experiments by P. Pillet {\it et al.\ } to study the photoassociation of
Cs$_2$ \cite{fioretti1999:ejd,comparat1999:jms,drag2000:ieee}.

Modeling the trap depletion dynamics as a function of photoassociation time
allowed us to extract precise absolute photoassociation rate coefficients,
i.~e.\ with an accuracy of better than a factor of 2 \cite{prodan2003:prl}.
We have observed photoassociation resonances down to the lowest vibrational
levels of the $0_g^-(P_{3/2})$ outer well. These resonances have very small
rate coefficients and could only be detected in previous measurements, because
they decay into bound molecules and are therefore observed by multiphoton
ionization. Our measurement yield the same quality as the ion spectra of
Ref.\ \cite{fioretti1999:ejd} without introducing the radiative transition to
bound molecules and the ionization efficiency into the signal strength. This
allows us to extract the Franck-Condon modulation of the initial
photoassociation step for all vibrational levels down to the $v=0$ level of
the $0_g^-(P_{3/2})$ external well.


\section{Experimental setup}

\begin{figure}
\includegraphics[width=8cm]{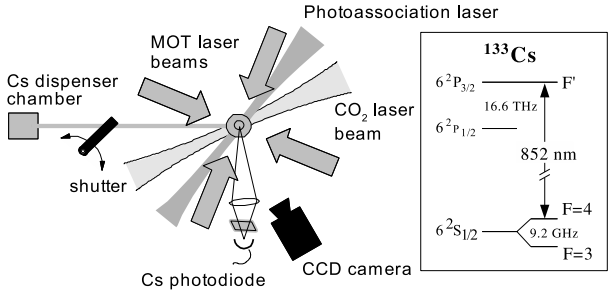}
\caption{\label{setup:fig} Experimental setup consisting of a cesium
  magneto-optical trap overlapped with a CO$_2$ optical dipole trap and a
  photoassociation laser.}
\end{figure}

The photoassociation experiments have been performed with an ensemble of
cesium atoms trapped in an optical dipole trap that is formed by the focus of
a CO$_2$ laser. As described in detail in Refs.\
\cite{engler2000:pra,mosk2001:apb}, the optical dipole trap is loaded from a
cesium magneto-optical trap which is superimposed in the focus of the CO$_2$
laser (see Fig.\ \ref{setup:fig}). The MOT is operated in five-beam
configuration and loads from a Zeeman slowed beam up to $10^8$ particles at a
density of $10^9\,\mbox{cm}^{-3}$ as inferred from absorption imaging.

After turning off the magnetic field of the MOT a brief molasses cooling phase
transfers about $5\times10^5$ atoms at a density of
$5\times10^{11}\,\mbox{cm}^{-3}$ and a temperature of 40\,$\mu$K into the
optical dipole trap \cite{engler2000:pra,mosk2001:apb}. The repumping beam is
extinguished several milliseconds before the main MOT beam, which pumps all Cs
atoms into the $F=3$ ground state. For the cold cesium atoms the CO$_2$ laser
focus with its potential depth of 0.8\,mK represents a harmonic trap with
axial and radial trap frequencies of $\omega_{ax}=12.8$\,Hz and
$\omega_{rad}=625$\,Hz respectively. In thermal equilibrium the density
distribution of the atoms in the trap is given by a cylindrically symmetric
Gaussian distribution with standard deviations of 600\,$\mu$m and 13\,$\mu$m
in the axial and radial directions, respectively. The lifetime of the atoms in
the dipole trap is of the order of 100\,s, due to collisions with residual gas
atoms.

The photoassociation light is provided by a widely tunable Titanium:Sapphire
laser (Coherent MBR 110) system, which delivers a typical output power of
200\,mW with a line width of about 100\,kHz, after passing through an optical
isolator and a single-mode optical fibre. Relative frequency changes are
monitored by measuring the transmission signal of a fraction of the
Ti:Sapphire laser beam through a confocal cavity with a free spectral range of
1000\,MHz. In order to decrease the frequency spacing of the transmission
peaks, two side-bands are modulated onto the cavity path of the Ti:Sapphire
beam at $\pm$200\,MHz using a double-pass AOM setup. The length of the cavity
is stabilized by locking one of the cavity mirrors with a piezo actuator to
the fringe of the transmission signal of a superimposed cesium-spectroscopy
stabilized diode laser. This provides a relative frequency accuracy of about
5\,MHz. The absolute laser frequency is measured with a commercial wavemeter
(Burleigh WA 1000) with an accuracy of 500\,MHz. The photoassociation laser is
passed through the trapped cesium cloud in the focus of the CO$_2$ laser at an
angle of $\theta=22.5^\circ$ with respect to the CO$_2$ laser beam. The width
of the Ti:Sapphire beam at the trap center amounts to 150\,$\mu$m. The
intensity of the photoassociation beam amounted to typically 50\,W/cm$^2$ and
was increased for the weakest vibrational transitions up to 300\,W/cm$^2$.

Once the cesium atoms are loaded into the dipole trap, the shutter of the
photoassociation laser is opened and the atom cloud is illuminated for up to
1000\,ms. For some data sets a short molasses cooling pulse is applied after
half of the photoassociation time to lower the temperature and thereby
increase the photoassociation rate. Finally the CO$_2$ laser light is
extinguished and all remaining cesium atoms are recaptured into the
magneto-optical trap. The number of recaptured cesium atoms is obtained from
the fluorescence signal of the MOT with a relative accuracy of better than
5\,\%. The fluorescence signal is calibrated to the absolute atom number using
absorption images, which leads to an absolute accuracy of the atom number of
30\,\%. Due to the very long storage time, the photoassociation-laser induced
atom loss signal is a precise measure for the formation of cold molecules
through photoassociation. To measure photoassociation spectra the frequency of
the Ti:Sapphire laser is scanned and the loading, photoassociation and
recapture cycle is repeated. The frequency range scanned during one cycle is
3\,MHz.

\begin{figure}
\includegraphics[width=8cm]{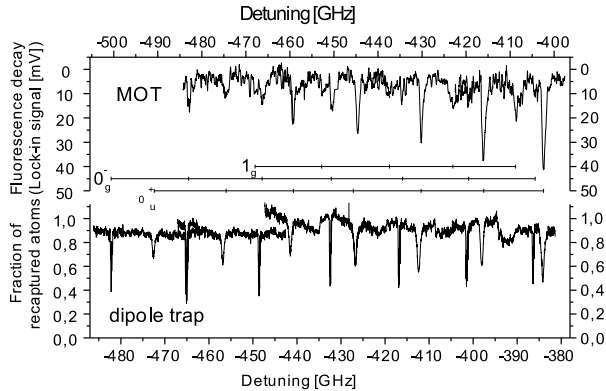}
\caption{\label{fig:mot_co2} Comparison of a wide-scan photoassociation
  spectrum measured in the far-detuned optical dipole trap (lower trace) with
  the same spectrum obtained in the magneto-optical trap (MOT) shown in the
  upper trace.}
\end{figure}

A wide photoassociation spectrum, composed of several individual scans, is
shown in the lower trace of Fig.\ \ref{fig:mot_co2}.  It is compared to a
photoassociation scan obtained directly in the magneto-optical trap using a
lock-in scheme to measure the fluorescence reduction that occurs when a
photoassociation resonance adds a loss channel to the MOT (upper trace). The
spectra show vibrational progressions in the 0$_g^-$, 0$_u^+$ and 1$_g$
molecular states that correlate to the $6s+6p_{3/2}$ asymptote.  Note the
exquisite stability of trap loading (overall fluctuations in particle number
$<5\%$). This comparison shows a significant improvement of the
signal-to-noise ratio for the optical dipole trap measurement. Furthermore,
the Cs hyperfine state in the MOT is $F=4$, whereas in the dipole trap it was
chosen to be the absolute ground state, $F=3$. This is reflected in the
different amplitudes of the resonances in the MOT compared to the dipole trap.


\section{Photoassociation results}


\subsection{Photoassociation into the $0_g^-(P_{3/2})$ outer well}

\begin{figure}
\includegraphics[width=8cm]{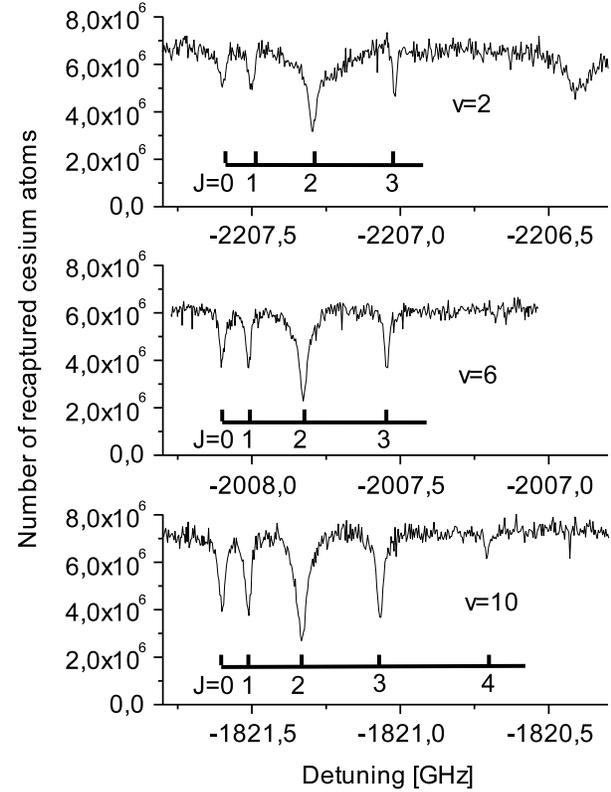}
\caption{\label{fig:spectra} Photoassociation trap loss spectra as a function
  of the photoassociation laser detuning relative to the $6s$
  $^2S_{1/2}$($F=4$)$\rightarrow$ $6p$ $^2P_{3/2}$($F'=5$) asymptote for v=2,
  6 and 10 in the 0$_g^-$ outer well. Rotational progressions are observed for
  each vibrational level up to J=3. For $v=10$ also a weak J=4 transition is
  observed. The resonance at -2208 GHz detuning in the upper panel is
  attributed to another electronic state.}
\end{figure}

By measuring photoassociation induced trap loss from the optical dipole trap,
the vibrational progression of the 0$_g^-(P_{3/2})$ outer well of Cs$_2$ (see
Fig.\ \ref{fig:photoassociation}) has been observed. Selected photoassociation
(PA) spectra are shown in Fig.\ \ref{fig:spectra} for the vibrational levels
$v=2$, 6 and 10. It can be seen that the number of recaptured atoms drops
significantly when the PA laser is tuned to a resonance with a rovibrational
state, leading to a strong spectroscopic signature. Outside of a resonance
essentially 100\,\% of the atoms transferred to the dipole trap are
recaptured. The noise on the baseline corresponds to the shot-to-shot
fluctuations in the number of atoms transferred to the dipole trap.

In the 0$_g^-$ state narrow resonances are observed, showing a rotational
progression for each vibrational level. Vibrational quantum numbers are
assigned to each spectrum based on the accurate spectroscopic investigation of
Ref.\ \cite{fioretti1999:ejd}.  From the rotational progression for each
vibrational level rotational constants are extracted, which are found to be in
good agreement with previous studies (see table \ref{table:rotation}).

It is interesting to note that we observe rigid rotor states of the cesium
molecule up to $J=4$.  For pure $s$-wave scattering angular momentum
conservation predicts a maximum rotational excitation of $J=2$ for optical
dipole excitation from the triplet ground state. This is indeed the strongest
rotational line in all the photoassociation spectra. Larger rotational states
can be rationalized due to additional angular momentum contributions, such as
from orbital angular momentum due to $p$-wave or $d$-wave scattering or from
the total nuclear spin of the atom pair. Actually, the height of the $p$-wave
centrifugal barrier is estimated to be 35\,$\mu$K, assuming that the initial
collisional channel proceeds via the $a^3\Sigma_u^+(6s+6s)$ state only. Given
the initial temperature of the cesium atoms of 40\,$\mu$K, this leads us to
attribute the $J=3$ resonances to $p$-wave collisions. During the
photoassociation time in the trap the temperature increases to about
60\,$\mu$K, as described in detail below. The origin of the weak $J=4$
component may therefore be a $d$-wave collision (barrier height about
200\,$\mu$K) of two fast atoms in the thermal distributions that either tunnel
through or surmount the barrier, even though both the $p$- and $d$-wave
barriers are found too weak to support a shape resonance that could increase
the tunneling probability. Light-induced mechanisms such as ``flux
enhancement'' which are important in magneto-optical traps
\cite{comparat1999:jms} do not play a role in a far-detuned optical dipole
trap.


\subsection{\label{trap_dynamics:sect}Model of the trap loss dynamics}

Owing to the controlled conditions that prevail inside the optical dipole trap
it is feasible to convert the trap loss signal into an absolute
photoassociation rate coefficient. For high photoassociation laser intensities
or long interaction times the photoassociation process can easily deplete more
than 50\,\% of the trapped atoms. Therefore a model is needed to describe the
depletion of the trapped atoms. This model assumes photoassociation from an
atomic cloud at thermal equilibrium and takes into account the time dependence
of both the atomic density and the temperature inside the optical dipole
trap. The basis of the depletion model is the rate equation for trap loss
\begin{equation}
\label{rate:eq}
\frac{dN}{dt} = -2 \int G(\vec x, T) \, n^2(\vec x, T) \, d^3x
\end{equation}
which describes the loss of two atoms per collision event and depends on the
photoassociation rate coefficient $G(\vec x,T)$ and the pair density $n^2(\vec
x, T)$. The latter is given by the square of the cylindrically symmetric
Gaussian density distribution of the thermal trapped atom cloud. The rate
coefficient $G(\vec x,T)$ needs special attention, because it depends on the
photoassociation laser intensity and therefore becomes position dependent if
the PA laser intensity varies spatially. In the experiment the width of the PA
laser of $w=150$\,$\mu$m is much larger than the radial extension of the
atomic cloud (13\,$\mu$m), but smaller than the axial extension
(800\,$\mu$m). The intensity therefore changes only along the axial direction,
denoted $z$, and has a Gaussian profile. In the low-intensity limit ($G
\propto$ intensity) one obtains
\begin{equation}
G(\vec x, T) = G_0(T) e^{- 2 \sin^2 \theta \, z^2 / w^2}.
\end{equation}
Note that the width of the distribution is given by the PA laser width and the
intersection angle of the PA laser with the trap axis $\theta$. The peak rate
coefficient $G_0(T)$ corresponds to the peak intensity $2 P / (\pi w^2)$ of
the photoassociation laser with power $P$.

Integration of Eq.\ (\ref{rate:eq}) yields a differential equation for the
particle number $N$,
\begin{equation}
\label{differential:eq}
\frac{dN}{dt} = -2 G_0(T) \frac{N^2}{V(T)},
\end{equation}
where the effective volume $V(T)$ depends on the PA laser and CO$_2$ trap
parameters and is given by
\begin{equation}
V(T) = \left ( \frac{2 \pi k_B T }{m_{Cs} \omega_{rad} \omega_{ax}} \right ) ^2
       \frac{ \sqrt{8} \sin \theta}{\sqrt{\pi} w}
     = V_0 T^2.
\end{equation}

The connection between particle number and temperature is taken from a
thermodynamic model of the trap depletion which was also successfully employed
to study hyperfine-changing Cs-Cs and Li-Cs collisions
\cite{mudrich2004:pra}. This model assumes that each colliding atom pair upon
leaving the trap removes a total energy of $q 6 k_B T$ where the parameter $q$
is obtained from a fit to the experimental data yielding $0.5 \pm 0.05$. As a
consequence of this relation, temperature and particle number are linked by
the scaling law
\begin{equation}
\label{n_t_link:eq}
\frac{T}{T_0} = \frac{N}{N_0}^{q-1}.
\end{equation}

Given the temperature dependence of the rate coefficient at low temperatures
$G_0(T) \propto T^{-\frac{1}{2}}$, Eq.\ (\ref{differential:eq}) is integrated
analytically yielding the number of trapped atoms and the temperature as a
function of the storage time and the photoassociation rate coefficient. This
is inverted to give the rate coefficient for the initial temperature $T_0$ as
a function of the remaining atoms $N(t_{PA})$ after a photoassociation time
$t_{PA}$,
\begin{equation}
\label{eq:pa_rate}
G_0(T_0) = \frac{V_0 T_0^2}{(9/2) N_0 t_{PA}} 
\left ( \left (N(t_{PA}) / N_0 \right )^{-9/4} - 1 \right ) .
\end{equation}
The cooling pulse that has been applied for some data sets after $t_{PA}/2$
was observed to reduce temperature and particle number of the trapped atoms by
a factor of $\alpha \sim 0.75$. It is accounted for by a piece-wise solution
of the differential equation (\ref{differential:eq}) and leads for these data
sets to a correction factor of $1/(1+\alpha^{3/2}) \sim 0.4$ to the rate
coefficient $G_0$.

\begin{figure}
\includegraphics[width=8cm]{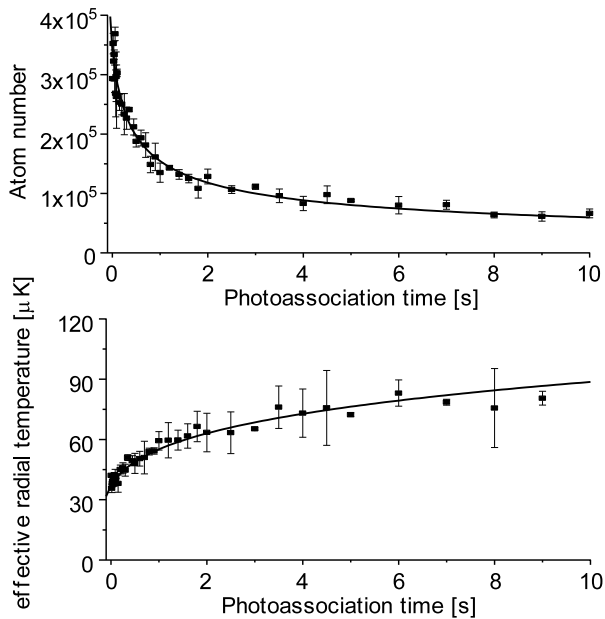}
\caption{\label{fig:traploss} Upper panel: number of atoms remaining in the
  trap as a function of the photoassociation laser interaction time; the solid
  line is a fit to the trap depletion model from which the absolute
  photoassociation rate coefficient is obtained. Lower panel: translational
  temperature of the remaining atoms as a function of photoassociation time,
  measured by absorption imaging of the radial expansion after release of the
  atoms from the CO$_2$ trap; the solid line is the fit of the trap
  depletion model with the initial temperature as the only free parameter.  
  Data are shown for the $v=20$, $J=2$ photoassociation resonance in the
  0$_g^-$ outer well.}
\end{figure}

The trap depletion model was verified experimentally for a single
photoassociation resonance by measuring the atom number, the atomic density
and the temperature as a function of photoassociation time using absorption
imaging. The result in Fig.\ \ref{fig:traploss} shows the measured decrease of
the number of cesium atoms. As can be seen, the temperature of the atoms is
increasing from 40 to about 80\,$\mu$K during 10\,s of interaction time with
the photoassociation laser, which is due to the coldest atoms associating most
favorably. Within the photoassociation time of 1\,s that is employed to
measure photoassociation spectra, the temperature increases to about
60\,$\mu$K. This effect is included in the model through
Eq. (\ref{n_t_link:eq}). One can see that the agreement to our trap depletion
model is satisfactory, despite its simplicity.

Using this model, specifically Eq.\ \ref{eq:pa_rate}, absolute
photoassociation rate coefficients are calculated for the resonances of the
$v=2$ spectrum in Fig.\ \ref{fig:spectra} and values between
$5\cdot10^{-12}$\,cm$^3$/s for $J=0$ and $3\cdot10^{-11}$\,cm$^3$/s for
$J=2$ are obtained. The PA laser intensity amounted to 90\,W/cm$^2$. The
accuracy of the obtained rate coefficients is estimated to be about 40\,\%,
which is mainly due to the accuracy of the absolute particle number as
obtained from absorption imaging (estimated to be 30\,\%) and the accuracy of
the cesium temperature (estimated to be 10\,\%).

In Ref.\ \cite{drag2000:ieee} absolute photoassociation rate coefficients were
measured in a magneto-optical trap and compared to theoretical calculations.
The measurements were carried out for several PA resonances in the $0_u^+$ and
$0_g^-$ states at small detunings and rate coefficients of between
$1\cdot10^{-11}$\,cm$^3$/s and $5\cdot10^{-11}$\,cm$^3$/s were measured. In
the same paper theoretical rate coefficients are calculated using a
perturbative model in good agreement with the measured rates. For
photoassociation into the 0$_g^-$ state at large detunings, corresponding to
the low vibrational levels that are studied here, the model in Ref.\
\cite{drag2000:ieee} predicts rate coefficients of a few times
10$^{-12}$\,cm$^3$/s for a laser intensity of 55\,W/cm$^2$. Our measured rate
coefficients are somewhat larger than this range. This is partly accounted for
by the higher laser intensity by a factor of 1.5 and the lower temperature in
our experiment of almost a factor of 4 which together predicts higher rate
coefficient by about a factor of three or values of around
10$^{-11}$\,cm$^3$/s \cite{pillet1997:jpb}. This is in reasonable agreement
with the measured values. An even better test of the quality of the absolute
rate coefficient measurements in the dipole trap has been carried out in our
group by studying saturation of the photoassociation rate coefficient at the
unitarity limit \cite{kraft2004:pra}. These measurements agree within the
estimated 40\,\% accuracy with theoretical prediction for the unitarity
limited photoassociation rate coefficient.


\subsection{Franck-Condon modulation in the $0_g^-(P_{3/2})$ outer well}

The photoassociation efficiency depends crucially on the Franck-Condon (FC)
factor $\langle \psi(v)| \phi \rangle$ involving the continuum wavefunction $|
\phi \rangle$ of the initial collisional state and the molecular wavefunction
$| \psi(v) \rangle$ of the PA level in the excited molecular state. The
brackets hold for the integration over the interatomic coordinate $R$. In this
approach the generally $R$-dependent dipole transition moment is approximated
to be constant for the interatomic distances at which photoassociation occurs.
The FC factor is known to exhibit strong variations with the vibrational
quantum number $v$ of the excited state
\cite{thorsheim1987:prl,cote1995:prl,drag2000:prl}.  Indeed, the PA process
involves an excited potential curve and an initial potential curve varying as
$R^{-3}$ and $R^{-6}$ respectively at large distances. For small detunings of
the PA laser, the overlap integral is mainly determined by the value of the
initial continuum wavefunction at the external turning point of the excited
potential. The intensity pattern of the PA spectrum then reflects the
oscillations of the continuum wavefuntion. At larger detunings, radial wave
functions in both channels may locally oscillate with similar frequency,
inducing abrupt changes in the overlap integral.

In this work we present measurements of the Franck-Condon modulation directly
through the analysis of trap loss spectra for the lowest vibrational levels in
the 0$_g^-(P_{3/2})$ external well, in contrast to previous measurements
\cite{fioretti1999:ejd} which detected ground state molecules by
photoionization and therefore measured the FC modulation multiplied by the
radiative decay probability and the ionization efficiency. Furthermore, in the
present experiment the $F=3$ initial state is chosen for the cesium atoms, in
contrast to the measurements with $F=4$ atoms in a magneto-optical trap.

To connect the photoassociation rate coefficient with the Franck-Condon factor
we employ the theoretical description by Bohn and Julienne
\cite{bohn1999:pra}, which is, for low photoassociation intensities,
equivalent to the treatment proposed in several papers
\cite{julienne1996:nist,napolitano1994:prl,pillet1997:jpb}. It yields a
scattering probability
\begin{equation}
\label{eq:fcf1}
p_{sc} = \frac{G}{G_{\mbox{unitarity}}}
  = \frac{ \Gamma } { \gamma / 4 }.
\end{equation}
This relation holds if the PA rate coefficient $G$ is small compared to the
unitarity limited maximum rate PA coefficient $G_{\mbox{unitarity}} \propto
T^{-1/2}$. The scattering probability $p_{sc}$ thus depends on the coupling
constant $\Gamma$ and the natural line width $\gamma$ of the electronic
transition. The coupling constant $\Gamma$ is given by
\begin{equation}
\label{eq:fcf2}
\Gamma = 2 \pi (V_{eg})^2 
| \langle \psi(v) | \phi (E_{\rm kin} \rightarrow 0) \rangle |^2,
\end{equation}
where $(V_{eq})^2$ is the Rabi frequency squared of the electronic transition,
which is proportional to the laser intensity.

\begin{figure}
\includegraphics[width=8cm]{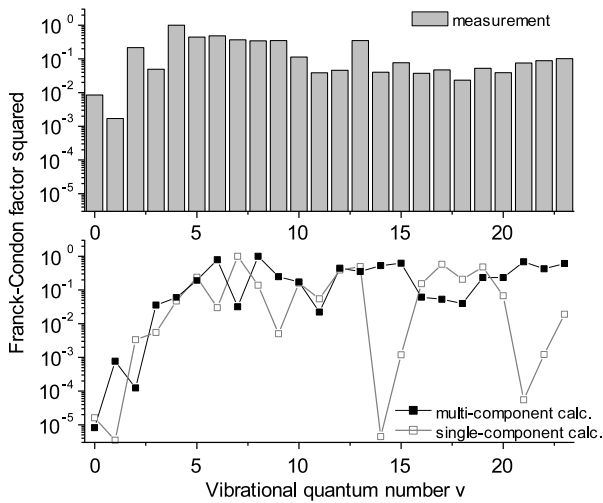}
\caption{\label{fig:franck_condon} Experimental data (upper panel) and
  theoretical results (lower panel) for the square of the Franck-Condon factor
  as a function of the vibrational level for photoassociation resonances into
  $J=2$ states in the 0$_g^-$ outer well. Each curve is normalized to the
  maximum Franck-Condon factor.
}
\end{figure}

For each $J=2$ rotational state of the vibrational levels $v=0$ to 23 we have
measured the photoassociation rate coefficient using the trap depletion model
described above. In the course of this work we investigated if more
vibrational levels could be found below the $v=0$ level, which would cause an
incorrect assignment of the vibrational progression. However, no additional
vibrational levels were found for larger detunings than the $v=0$ level.  For
each observed vibrational level the Franck-Condon (FC) factor is calculated
using Eqs.\ (\ref{eq:fcf1}) and (\ref{eq:fcf2}). The measurements for $v \leq
10$ were carried out with an intermediate cooling pulse and were corrected
accordingly (see section \ref{trap_dynamics:sect}). The accuracy of the
individual FC factors is estimated to be 60\,\%, mainly given by the accuracy
of the rate coefficient (see above) and the PA laser intensity. To remove the
energy-normalization of the continuum wavefunction $\phi$ all FC factors have
been normalized to the maximum FC factor, $v=4$, which now amounts to 1. The
result is shown in the upper panel of Fig.\ \ref{fig:franck_condon}. The
measured FC factors show a clear step modulation starting with a very small
value for $v=0$ and 1 and increasing by a factor of 100 from $v=1$ to 2. For
higher vibrational levels a much weaker modulation is observed with a local
maximum near $v=4$ and a local minimum around $v=18$.

Theoretical modeling of the Franck-Condon factor modulation has been carried
out for comparison with the experimental results. This requires the
computation of the $| \psi(v) \rangle$ and $| \phi \rangle$ radial wave
functions. The vibrational wavefunctions of the $0_g^-(P_{3/2})$ external well
are easily obtained from a standard Numerov integration using the
Rydberg-Klein-Rees (RKR) potential of Ref.\ \cite{fioretti1999:ejd}. In
principle, the main task is the integration of the continuum wave
function. Only when the atoms are prepared in the fully stretched states
$F=4,M_F=4$, the collisional entrance channel is fully correlated to a single
state, i.e. the lowest $a^3\Sigma_u^+$ state: it is described accurately
enough by an available theoretical potential curve, connected at large
distances to an accurate $\sum{C_n/R^n}$ asymptotic expansion (see for
instance Refs.\ \cite{amiot2002:jcp,vanhaecke2004:ejd}), and adapted to
reproduce the measured cesium triplet scattering length
\cite{kerman2001:ras}. This type of calculations has been performed in Ref.\
\cite{drag2000:prl} and is referred to in the following as the
single-component calculation.

In the present experiment the atoms are not polarized, so that many channels
with different internal hyperfine quantum numbers contribute to the
process. Moreover, each collisional channel is described by a radial
wavefunction which results from the coupling between components of both
$a^3\Sigma_u^+$ and $X^1\Sigma_g^+$ electronic states, interacting through the
hyperfine hamiltonian. Indeed the vibrational motion of the excited levels
presented here proceeds over the 15-30$a_0$ range, which is the recoupling
range for the hyperfine interactions in cesium. The complexity of this coupled
channel problem has lead us to perform the calculation for two colliding atoms
occupying the $F=3,M_F=3$ state. The total continuum wavefunction is then a
three-component molecular wavefunction characterized by the projection $M=6$
of the total angular momentum. With this multi-component wavefunction the
Franck-Condon modulation is calculated.

In the lower panel of Fig.\ \ref{fig:franck_condon} the Franck-Condon factors
are shown that are calculated using the multi-component wave function (full
squares) together with the result for the single-component wave function (open
squares). Again all values are normalized to the maximum FC factor. Both
calculations clearly show a strong increase for small vibrational levels
similar to the observation in the experiment. For the multi-component
calculation the steepest increase is observed between $v=2$ and 3 in contrast
to the experimentally found step between $v=1$ and 2, which is also found in
the single-component calculation. For larger vibrational levels the FC factors
of the multi-component calculation exhibit a somewhat larger amplitude
variation than the measured data, but the overall FC factors are fairly
constant. Even thought the atoms in the experiment are not polarized in
contrast to the calculation, a similar trend is observed in the experimental
data. A local minimum near $v=18$ is found in the multi-component calculation,
similar to the measurement, but other local minima near $v=7$ and 11 are not
found in the measurement. The single-component calculation shows much
deeper minima for $v=14$ and 21, which are not present in the more accurate
multi-component calculation. Thus one may speculate that also the minima in the
multi-component FC calculation may become filled in by angular momentum
components that are not included in the calculation but certainly present in
the experiment. To resolve this a full FC calculation or a measurement with
polarized atoms would be necessary.


\section{Conclusions}

As it was shown in Ref.\ \cite{miller1993:prl}, photoassociation experiments
benefit significantly from the utilization of a conservative optical trap, due
to the high pair density and the long interaction times. Furthermore the
internal hyperfine state of the trapped atoms is well defined. In this work we
present a thorough quantitative analysis of the photoassociation in a
far-detuned optical dipole trap. Using a thermodynamic trap depletion model
absolute photoassociation rate coefficients were obtained with an accuracy of
about 40\,\%.

The high sensitivity of photoassociation in a dipole trap was demonstrated by
studying photoassociation resonances into the lowest vibrational levels of the
outer well of the $0_g^-(P_{3/2})$ state through trap loss. From these
measurements the Franck-Condon factors for the initial single-photon
excitation step of the photoassociation process were extracted for the lowest
24 vibrational levels of the $0_g^-(P_{3/2})$ outer well and compared to a
theoretical calculation. Previously, these transitions could only be observed
by multiphoton ionization of ground state molecules. No further
photoassociation resonances at larger detuning than $v=0$ were found,
confirming the assignment in Ref.\ \cite{fioretti1999:ejd}.

In the current experiment ground state Cs$_2$ molecules are presumably already
produced and trapped, but not observed due to the lack of an appropriate
detection system.  In future experiments multiphoton ionization will be used
to detect and study these trapped molecules.


\section{Acknowledgments}

The experiments were performed while the group was still at the
Max-Planck-Institut f{\"ur} Kernphysik in Heidelberg and we wish to thank the
institute, in particular D. Schwalm, for generous support. This work is
supported by the Deutsche Forschungsgemeinschaft in the Schwerpunktprogramm
1116 ,,Interactions in Ultracold Atomic and Molecular Gases'' under
WE-2661/1-2. We also acknowledge support by the EU research training network
,,Cold molecules'' (COMOL), under the contract number HPRN-2002-00290.


\bibliographystyle{aip}
\bibliography{../../References/cs_pa,../../References/group,../../References/group_unp}

\begin{table}
\begin{tabular}{|c|c|c|c|c|c|}
\hline
v & $\delta$ [GHz] $^a$ & $\delta$ [GHz] $^b$ & 
    $B$ [MHz] $^a$ & $B$ [MHz] $^b$ \\
\hline
   2  &  -2207.6  &  -2207.6(0.5)  &  48.9  & 49.3(1.0) \\
   6  &  -2008.2  &  -2008.1(0.5)  &  46.8  & 45.6(1.0) \\
  10  &  -1821.6  &  -1821.6(0.5)  &  45.0  & 44.6(1.0) \\
\hline
\end{tabular}
\caption{\label{table:rotation}Detuning $\delta$ of the $J=0$ level relative
  to the $6s$ $^2S_{1/2}$($F=4$)$\rightarrow$ $6p$ $^2P_{3/2}$($F'=5$)
  asymptote and rotational constants $B$ for selected vibrational states of
  the $0_g^-(P_{3/2})$ ($^a$: Ref. \cite{fioretti1999:ejd}; $^b$: this work,
  estimated accuracies are denoted in brackets).}
\end{table}\

\end{document}